\documentclass[preprint,12pt]{elsarticle}
\usepackage{microtype}
\usepackage{lineno}  
\usepackage{url}  

\usepackage{amssymb}
\usepackage{braket}
\usepackage{epsfig}
\usepackage{epstopdf}
\usepackage{multirow}
\usepackage{booktabs}
\usepackage{graphicx}
\usepackage[tbtags]{amsmath}

\journal{New Astronomy}

\begin{document}

\begin{frontmatter}

\title{On the role of pairing correlations in calculation of $\beta$-decay half-lives within QRPA formalism}

\author{Jameel-Un Nabi, Mavra Ishfaq\footnote{Corresponding Author}}
\address{Faculty of Engineering Sciences,\\GIK
Institute of Engineering Sciences and Technology,\\ Topi 23640,
Khyber
Pakhtunkhwa, Pakistan\\
{jameel@giki.edu.pk, mavra.ishfaq34@gmail.com}}

\begin{abstract}
In this paper we show that the proton-neutron residual interaction can play
an important role in the reliability of calculated $\beta$-decay
half-lives. It may also improve the prediction power of the
quasiparticle random phase approximation (QRPA) model. We further
demonstrate that a reasonable choice of the particle-particle
(attractive) and particle-hole force (repulsive) parameters can
result in calculated half-lives in very good comparison with the
measured ones. Pairing gaps have affect on calculated
half-lives which we explore in this paper. We present our half-lives calculation using the proton-neutron QRPA (pn-QRPA) model
possessing a multi-shell single-particle deformed space including a
schematic interaction for some medium mass neutron-deficient nuclei
undergoing $\beta^{+}$/EC decay. Our study shows a better agreement
with the available experimental data as compared to former
calculations.
\end{abstract}


\end{frontmatter}


%
\section{Introduction}
The $\beta$-decay properties are useful tools to better understand
the overall picture of nuclear structure \cite{Fer34}. Research on
unstable nuclei reveals that $\beta$-decay plays a pivotal role
among decay channels \cite{Tan17}. In the field of nuclear
astrophysics $\beta$-decay properties of neutron deficient nuclei
are involved in astrophysical $rp$-process and are required as input
parameters for running numerical simulations \cite{Sar05}.

The electron capture (EC) and $\beta^{+}$ decay (at times also referred to as positron decay) is
a very important decay mode for neutron deficient nuclei. Various
nuclear models have been used in the past to study the properties of
$\beta$-decay. Of special mention are those calculations based on
gross theory [e.g. \cite{Tak73}], quasiparticle random phase approximation
(QRPA) approaches [e.g.
\cite{Sta90,Hir93,Nab99,Wan16,Ni14,Nik05,Mar07,Lia08,Niu13,Tan17}]
and shell model [e.g. \cite{Mar99}]. The gross theory adopts a statistical
approach to get an estimate of the $\beta$-decay properties. On the
other hand, the shell model and QRPA approaches are microscopic in
nature. Usage of shell model has the obvious constraint of number of
basis states which creeps in as soon as we start to study heavy
nuclei. Shell model results may be accurate only for light cases
\cite{Tan17}. The QRPA approach leads to precise and schematic
information about $\beta$-decay properties \cite{Wan16}. Using this
method one can reproduce  available experimental data for
$\beta$-decay half-lives in a reliable and efficient manner.
Previously the finite-range droplet model (FRDM) and folded-Yukawa
single-particle potential was used to study nuclear properties of
around 9000 nuclei ranging from $^{16}$O to $^{339}$136 using the
QRPA approach \cite{Mol97}. In past research it was reported that
$\beta$-decay half-lives were significantly affected by the
proton-neutron pairing interaction \cite{Pan96,Wan16,Tan17}.

In this paper we calculate and analyze the half-lives of $\beta^{+}$/EC
decay for even-even medium mass neutron deficient nuclei with atomic
number in the range $Z = 24 - 34$ far from $\beta$-stability line.
We perform our calculation using a  proton-neutron QRPA (pn-QRPA)
base model in a multi-shell single-particle deformed space with
schematic and separable Gamow-Teller (GT) potential. We show that a
reasonable choice of the particle-particle ($pp$) and particle-hole
($ph$) GT strength parameters (see next section for further
explanation) may lead to improved prediction power and reliable
calculation of $\beta^{+}$/EC decay half-lives. The calculated
half-lives are later compared with measured data \cite{Aud17}, FRDM
\cite{Mol97} and the recent extended QRPA (EQRPA)
calculations \cite{Tan17}.

This paper is drafted in following manner. In Sec. 2 we explain our
base model (pn-QRPA) and admit necessary formalism for half-life
calculation within the framework of deformed pn-QRPA approach.
Results and discussion are presented in Section 3.
Conclusions and summary are given in Section 4.

\section{FORMALISM}
The role of residual interaction becomes important as the number of nucleons outside the closed major shell increases. Of special mention is the short-range part of the residual interaction which manifest itself as pairing correlations between the nucleons. This phenomenon is commonly referred to as nucleon pairing. This correlation lowers their total energy by an amount $2\Delta$ (where $\Delta$ is the pairing gap).

For our present study we employ the pn-QRPA model in a deformed
space using the  Nilsson+BCS formalism. We further incorporate
multi-shell single-particle states and include a schematic
interaction \cite{Sta90,Hir93}. See also Ref. \cite{Mut92}. Using
this base model we calculated $\beta$-decay half-lives for neutron deficient
even-even nuclei possessing neutron number
$N$ in the range of 18 to 36.

We started with a spherical basis $(a^{\dagger}_{jk}, a_{jk})$,
possessing $j$ as its total angular momentum and $k$ as the
associated z-component. This basis was later transformed to a
deformed (axial-symmetric) basis $(d^{\dagger}_{k\alpha},
d_{k\alpha})$ using the transformation equation

\textbf{\begin{equation}\label{df}
d^{\dagger}_{k\alpha}=\Sigma_{j}K^{k\alpha}_{j}a^{\dagger}_{jk}.
\end{equation}}

In Eq.~\ref{df} $K$ stands for matrix of transformation obtained
from set of Nilsson eigenfunctions, and $\alpha$ (represents additional quantum except $k$) which specify the Nilsson eigenstates. We employed the BCS calculation for neutron and
proton systems independently. We took a constant pairing force
possessing strength of $V$($V_n$, $V_p$ for neutrons and protons,
respectively),

\begin{equation}\label{pr}
\begin{split}
F^{pair}=-V\sum_{jkj^{'}k^{'}}(-1)^{l+j-k}a^{\dagger}_{jk}a^{\dagger}_{j-k} (-1)^{l^{'}+j^{'}-k^{'}}\\
a_{j^{'}-k}a_{j^{'}k^{'}},
\end{split}
\end{equation}
where the sum over $k$ and $k^{'}$ was limited to $k$, $k^{'}$ $>$ 0
and $l$ stands for orbital angular momentum. The BCS calculation
provides occupation amplitudes $u_{k\alpha}$, $v_{k\alpha}$ (which
fulfills $u^{2}_{k\alpha}$+$v^{2}_{k\alpha}$=1) and quasiparticle
(q.p.) energies $\varepsilon_{k\alpha}$. It is to be noted that the pairing gaps ($\Delta$) were determined empirically in our calculation and is discussed towards the end of  this section.  Later we added a q.p. basis
$(c^{\dagger}_{k\alpha}, c_{k\alpha})$ by introducing a Bogoliubov
transformation \textbf{\begin{equation}\label{qbas}
c^{\dagger}_{k\alpha}=u_{k\alpha}d^{\dagger}_{k\alpha}-v_{k\alpha}d_{\bar{k}\alpha}\\
c^{\dagger}_{\bar{k}\alpha}=u_{k\alpha}d^{\dagger}_{\bar{k}\alpha}+v_{k\alpha}d_{k\alpha}.
\end{equation}}
Here $\bar{k}$ is the time inverted state of $k$ and $c/c^{\dagger}$
stands for the q.p. annihilation/creation operators which eventually
entered our RPA equation. Creation operators of QRPA phonons was
introduced using the relation
\begin{equation}\label{co}
C^{\dagger}_{\omega}(\mu)=\sum_{pn}[X^{pn}_{\omega}(\mu)c^{\dagger}_{p}c^{\dagger}_{n}-Y^{pn}_{\omega}(\mu)c_{n}c_{\overline{p}}].
\end{equation}
Indices $n$ and $p$ in Eq.~\ref{co} stand for $m_{n}\alpha_{n}$ and
$m_{p}\alpha_{p}$, respectively, and distinguish between neutron and
proton single-(quasi)-particle states. $X^{pn}_{\omega}$ and
$Y^{pn}_{\omega}$ are amplitudes for forward and backward-going,
respectively. They are the eigenfunctions for the RPA matrix
equation. $\omega$ represents the corresponding energy eigenvalues
of the eigenstates.

We consider the $ph$ Gamow-Teller (GT) force for RPA calculation using the relation

\textbf{\begin{equation}\label{ph}
F^{ph}= +2\chi\sum^{1}_{\mu= -1}(-1)^{\mu}Y_{\mu}Y^{\dagger}_{-\mu}\\
\end{equation}
\begin{equation}\label{ph1}
Y_{\mu}= \sum_{j_{p}m_{p}j_{n}m_{n}}<j_{p}m_{p}\mid
\tau_{-}\sigma_{\mu}\mid
j_{n}m_{n}>c^{\dagger}_{j_{p}m_{p}}c_{j_{n}m_{n}}.
\end{equation}}

The corresponding $pp$ GT force was calculated using

\textbf{\begin{equation}\label{ph} F^{pp}= -2\kappa\sum^{1}_{\mu=
-1}(-1)^{\mu}Z_{\mu}^{\dagger}Z_{-\mu}.
\end{equation}}
with
\begin{equation} \label{GrindEQ13_}
\begin{split}
     Z_{\mu }^{+} =\sum _{j_{p} m_{p} j_{n} m_{n}
}<j_{n} m_{n} |(\tau_{-} \sigma _{\mu } )^{+} |j_{p} m_{p}
> \\
(-1)^{l_{n} +j_{n} -m_{n} } c_{j_{p} m_{p} }^{+} c_{j_{n} -m_{n}
}^{+},
\end{split}
\end{equation}
In Eqs.~\ref{ph1} and ~\ref{GrindEQ13_},  $\tau_{\pm}$ is the
isospin raising (lowering) operator. The $\tau_+$ ($\tau_-$)
operator transforms a proton (neutron) to a neutron (proton),
$\sigma$ is the Pauli matrix. The $pp$ and $ph$ forces were characterized
by interaction constants $\kappa$ and $\chi$, respectively. All
remaining symbols have their usual meanings.

The construction of low-lying excited states for odd-A nuclei is done using the following recipe in our model: \\
(1) By exciting odd numbers of neutron from the ground state to high
(excited state) energy states.\\
(2) By three-neutron states, corresponding to excitation of a
neutron,\\
(3) By excitation of a proton, corresponding to  one neutron and two protons states.\\
The quasiparticle states in an even proton and odd neutron nucleus
can be obtained from Eqs.~\ref{qpr} and~\ref{nqpr} by the
interchange of neutron and proton states
($\eta\longleftrightarrow \pi$) and the annihilation (creation)
operators of QRPA phonons ($A^+_{\omega}(-\mu)\longleftrightarrow A^+_{\omega}(\mu)$).\\
The quasiparticles transition from lower to higher energy levels
for odd-proton and even-neutron state is attainable.

\begin{equation}\label{qpr}
\begin{split}
\mid{\pi_1\pi_2\pi_{3corr}}\rangle=\alpha^+_{\pi_1}\alpha^+_{\pi_2}\alpha^+_{\pi_3}|{-}\rangle+\\ \dfrac{1}{2}\sum\limits_{\pi'_1\pi'_2\eta'\omega}\alpha^+_{\pi'_1}\alpha^+_{\pi'_2}\alpha^+_{\eta'}A^+_{\omega}(\mu)|{-}\rangle\\
\langle{-}|[\alpha^+_{\pi'_1}\alpha^+_{\pi'_2}\alpha^+_{\eta'}A^+_{\omega}(\mu)]^+H\\
\alpha^+_{\pi_1}\alpha^+_{\pi_2}\alpha^+_{\pi3}|{-}\rangle
E_{\pi_1\pi_2\pi_3}(\pi'_1\pi'_2\eta',\omega),
\end{split}
\end{equation}
here '$H$' stands for the Hamiltonian obtained from separable
particle-hole (ph) and particle-particle (pp) forces by the
Bogoliubov transformation playing vital role for phonon-quasiparticle coupling. The sum is taken over all phonons and quasiparticle proton (neutron) states, satisfying $m_{\pi}-m_{\eta}=\mu$.
The symbols $\pi$ and $\eta$ indicate proton and neutron states,
respectively.
\begin{equation}\label{nqpr}
\begin{split}
|{\pi_1\eta_1\eta_{2corr}}\rangle=\alpha^+_{\pi_1}\alpha^+_{\eta_1}\alpha^+_{\eta_2}|{-}\rangle+ \\ \dfrac{1}{2}\sum\limits_{\pi'_1\pi'_2\eta'\omega}\alpha^+_{\pi'_1}\alpha^+_{\pi'_2}\alpha^+_{\eta'}A^+_{\omega}(-\mu)|{-}\rangle\\
\langle{-}|[\alpha^+_{\pi'_1}\alpha^+_{\pi'_2}\alpha^+_{\eta'}A^+_{\omega}(-\mu)]^+ \\ H \alpha^+_{\pi_1}\alpha^+_{\eta_1}\alpha^+_{\eta_2}|{-}\rangle E_{\pi_1\eta_1\eta_2}(\pi'_1\pi'_2\eta',\omega)+\dfrac{1}{6}\\
\sum\limits_{\eta'_1\eta'_2\eta'_3\omega}\alpha^+_{\eta'_1}\alpha^+_{\eta'_2}\alpha^+_{\eta'_3}A^+_{\omega}(\mu)\\|{-}\rangle \langle{-}|[\alpha^+_{\eta'_1}\alpha^+_{\eta'_2}\alpha^+_{\eta'_3}A^+_{\omega}(\mu)]^+\\
H\alpha^+_{\pi_1}\alpha^+_{\eta_1}\alpha^+_{\eta_2}|{-}\rangle
E_{\pi_1\eta_1\eta_2}(\eta'_1\eta'_2\eta'_3,\omega),
\end{split}
\end{equation}
with the energy denominator of first order perturbation,
\begin{equation}
E_{abcd}(def,\omega)=\dfrac{1}{(\epsilon_a+\epsilon_b+\epsilon_c)-(\epsilon_d+\epsilon_e+\epsilon_f+\omega)}.
\end{equation}

The formulae for multi quasi particle transitions and their
reduction to correlated one quasi particle state are given as:
\begin{equation}
\begin{split}
\langle{\pi^f_1\pi^f_2\eta^f_{1corr}}|t_{\pm}\sigma_{-\mu}|{\pi^i_1\pi^i_2\pi^i_{3corr}}\rangle=\\ \delta(\pi^f_1,\pi^i_2)\delta(\pi^f_2,\pi^i_3)\langle {\eta^f_{1corr}}|t_{\pm}\sigma_{-\mu}|{\pi^i_{1corr}}\rangle\\
-\delta(\pi^f_1,\pi^i_1)\delta(\pi^f_2,\pi^i_3)\langle{\eta^f_{\eta1}}|t_{\pm}\sigma_{-\mu}|{\pi^i_{2corr}}\rangle\\
+\delta(\pi^f_1,\pi^i_1)\delta(\pi^f_2,\pi^i_2)\langle{\eta^f_{\eta1}}|t_{\pm}\sigma_{-\mu}|{\pi^i_{3corr}}\rangle,
\end{split}
\end{equation}

\begin{equation}
\begin{split}
\langle{\pi^f_1\pi^f_2\eta^f_{1corr}}|t_{\pm}\sigma_{\mu}|{\pi^i_1\pi^i_2\pi^i_{2corr}}\rangle= \\ \delta(\eta^f_1,\eta^i_2)[\delta(\pi^f_1,\pi^i_1)\langle{\pi^f_{2corr}}|t_{\pm}\sigma_{\mu}|{\eta^i_{1corr}}\rangle\\
-\delta(\pi^f_2,\pi^i_1)\langle{\pi^f_{corr1}}|t_{\pm}\sigma_{\mu}|{\eta^i_{1corr}}\rangle]\\
-\delta(\eta^f_1,\eta^i_1)[\delta(\pi^f_1,\pi^i_1)\langle{\pi^f_{2corr}}|t_{\pm}\sigma_{\mu}|{\eta^i_{2corr}}\rangle\\
-\delta(\pi^f_2,\pi^i_1)\langle{\pi^f_{1corr}}|t_{\pm}\sigma_{\mu}|{\eta^i_{2corr}}\rangle],
\end{split}
\end{equation}

\begin{equation}
\begin{split}
\langle{\eta^f_1\eta^f_2\eta^f_{3corr}}|t_{\pm}\sigma_{-\mu}|{\pi^i_1\eta^i_1\eta^i_{2corr}}\rangle= \\ \delta(\eta^f_2,\eta^i_1)\delta(\eta^f_3,\eta^i_2)\langle{\eta^f_{1corr}}|t_{\pm}\sigma_{-\mu}|{\pi^i_{1corr}}\rangle\\
-\delta(\eta^f_1,\eta^i_1)\delta(\eta^f_3,\eta^i_2)\langle{\eta^f_{2corr}}|t_{\pm}\sigma_{-\mu}|{\pi^i_{1corr}}\rangle\\
+\delta(\eta^f_1,\eta^i_1)\delta(\eta^f_2,\eta^i_2)\langle{\eta^f_{3corr}}|t_{\pm}\sigma_{-\mu}|{\pi^i_{1corr}}\rangle.
\end{split}
\end{equation}

For odd-odd nuclei the quasiparticle transformation is expressed by
using two quasiparticle states of proton and neutron pair states
and four quasiparticle states of two nucleons. Similarly in this case,
transition amplitudes for the quasiparticle states are reduced into
the correlated one quasiparticle state and given as

\begin{equation}
\begin{split}
\langle{\pi^f_1\pi^f_{2corr}}|t_{\pm}\sigma_{\mu}|{\pi^i\eta^{i}_{corr}}\rangle=
\delta(\pi^f_1,\pi^i)\langle{\pi^f_{2{corr}}}|t_{\pm}\sigma_{\mu}|{\eta^i_{corr}}\rangle
\\-\delta(\pi^f_2,\pi^i)\langle{\pi^f_{1{corr}}}|t_{\pm}\sigma_{\mu}|{\eta^i_{corr}}\rangle,
\end{split}
\end{equation}

\begin{equation}
\begin{split}
\langle{\eta^f_1\eta^f_{2corr}}|t_{\pm}\sigma_{-\mu}|{\pi^in^i_{corr}}\rangle=
\delta(\eta^f_2,\eta^i)\langle{\eta^f_{1corr}}|t_{\pm}\sigma_{-\mu}|{\pi^i_{corr}}\rangle\\
-\delta(\eta^f_1,\eta^i)\langle{\eta^f_{2corr}}|t_{\pm}\sigma_{-\mu}|{\pi^i_{corr}}\rangle.
\end{split}
\end{equation}
The four quasiparticle states are simplified as following

\begin{equation}
\begin{split}
\langle{\pi^f_1\pi^f_2\eta^f_1\eta^f_{2corr}}|t_{\pm}\sigma_{-\mu}|{\pi^i_1\pi^i_2\pi^i_3\eta^i_{1corr}}\rangle=\\
\delta(\eta^f_2,\eta^i_1)[\delta(\pi^f_1,\pi^i_2)\delta(\pi^f_2,\pi^i_3)\langle{\eta^f_{1corr}}|t_{\pm}\sigma_{-\mu}|{\pi^i_{1corr}}\rangle\\
-\delta(\pi^f_1,\pi^i_1)\delta(\pi^f_2,\pi^i_3)\langle{\eta^f_{1corr}}|t_{\pm}\sigma_{-\mu}|{\pi^i_{2corr}}\rangle\\ +\delta(\pi^f_1,\pi^i_1)\delta(\pi^f_2,\pi^i_2)\langle{\eta^f_{1corr}}|t_{\pm}\sigma_{-\mu}|{\pi^i_{3corr}}\rangle]\\
-\delta(\eta^f_1,\eta^i_1)[\delta(\pi^f_1,\pi^i_2)\delta(\pi^f_2,\pi^i_3)\langle{\eta^f_{2corr}}|t_{\pm}\sigma_{-\mu}|{\pi^i_{1corr}}\rangle \\ -\delta(\pi^f_1,\pi^i_1)\delta(\pi^f_2,\pi^i_3)\langle{\eta^f_{2corr}}|t_{\pm}\sigma_{-\mu}|{\pi^i_{2corr}}\rangle\\
+\delta(\pi^f_1,\pi^i_1)\delta(\pi^f_2,\pi^i_2)\langle{\eta^f_{2corr}}|t_{\pm}\sigma_{-\mu}|{\pi^i_{3corr}}\rangle],
\end{split}
\end{equation}

\begin{equation}
\begin{split}
\langle{\pi^f_1\pi^f_2\pi^f_3\pi^f_{4corr}}|t_{\pm}\sigma_{\mu}|{\pi^i_1\pi^i_2\pi^i_3\eta^i_{1corr}}\rangle=\\ -\delta(\pi^f_2,\pi^i_1)\delta(\pi^f_3,\pi^i_2)\delta(\pi^f_4,\pi^i_3)\langle{\pi^f_{1corr}}|t_{\pm}\sigma_{\mu}|{\eta^i_{1corr}}\rangle\\
+\delta(\pi^f_1,\pi^i_1)\delta(\pi^f_3,\pi^i_2)\delta(\pi^f_4,\pi^i_3)\langle{\pi^f_{2corr}}|t_{\pm}\sigma_{\mu}|{\eta^i_{1corr}}\rangle\\ -\delta(\pi^f_1,\pi^i_1)\delta(\pi^f_2,\pi^i_2)\delta(\pi^f_4,\pi^i_3)\langle{\pi^f_{3corr}}|t_{\pm}\sigma_{\mu}|{\eta^i_{1corr}}\rangle\\
+\delta(\pi^f_1,\pi^i_1)\delta(\pi^f_2,\pi^i_2)\delta(\pi^f_3,\pi^i_3)\langle{\pi^f_{4corr}}|t_{\pm}\sigma_{\mu}|{\pi^i_{1corr}}\rangle,
\end{split}
\end{equation}

\begin{equation}
\begin{split}
\langle{\pi^f_1\pi^f_2\eta^f_1\eta^f_{2corr}}|t_{\pm}\sigma_{\mu}|{\pi^i_1\eta^i_1\eta^i_2\eta^i_{3corr}}\rangle=\\ \delta(\pi^f_1,\pi^i_1)[\delta(\eta^f_1,\eta^i_2)\delta(\eta^f_2,\eta^i_3)\langle{\pi^f_{2corr}}|t_{\pm}\sigma_{\mu}|{\eta^i_{1corr}}\rangle\\
-\delta(\eta^f_1,\eta^i_1)\delta(\eta^f_2,\eta^i_3)\langle{\pi^f_{2corr}}|t_{\pm}\sigma_{\mu}|{\eta^i_{2corr}}\rangle\\ +\delta(\eta^f_1,\eta^i_1)\delta(\eta^f_2,\eta^i_2)\langle{\pi^f_{2corr}}|t_{\pm}\sigma_{\mu}|{\eta^i_{3corr}}\rangle]\\
-\delta(\pi^f_2,\pi^i_1)[\delta(\eta^f_1,\eta^i_2)\delta(\eta^f_2,\eta^i_3)\langle{\pi^f_{1corr}}|t_{\pm}\sigma_{\mu}|{\eta^i_{1corr}}\rangle\\ -\delta(\eta^f_1,\eta^i_1)\delta(\eta^f_2,\eta^i_3)\langle{\pi^f_{1corr}}|t_{\pm}\sigma_{\mu}|{\eta^i_{2corr}}\rangle\\
+\delta(\eta^f_1,\eta^i_1)\delta(\eta^f_2,\eta^i_2)\langle{\pi^f_{1corr}}|t_{\pm}\sigma_{\mu}|{\eta^i_{3corr}}\rangle],
\end{split}
\end{equation}

\begin{equation}
\begin{split}
\langle{\eta^f_1\eta^f_2\eta^f_3\eta^f_{4corr}}|t_{\pm}\sigma_{-\mu}|{\pi^i_1\eta^i_1\eta^i_2\eta^i_{3corr}}\rangle=\\ +\delta(\eta^f_2,\pi^i_1)\delta(\eta^f_3,\eta^i_2)\delta(\eta^f_4,\eta^i_3)\langle{\eta^f_{1corr}}|t_{-\pm}\sigma_{\mu}|{\pi^i_{1corr}}\rangle\\
-\delta(\eta^f_1,\eta^i_1)\delta(\eta^f_3,\eta^i_2)\delta(\eta^f_4,\eta^i_3)\langle{\eta^f_{2corr}}|t_{\pm}\sigma_{-\mu}|{\pi^i_{1corr}}\rangle\\
+\delta(\eta^f_1,\eta^i_1)\delta(\eta^f_2,\eta^i_2)\delta(\eta^f_4,\eta^i_3)\langle{\eta^f_{3corr}}|t_{\pm}\sigma_{-\mu}|{\pi^i_{1corr}}\rangle\\
-\delta(\eta^f_1,\eta^i_1)\delta(\eta^f_2,\eta^i_2)\delta(\eta^f_3,\eta^i_3)\langle{\eta^f_{4corr}}|t_{\pm}\sigma_{-\mu}|{\pi^i_{1corr}}\rangle.
\end{split}
\end{equation}
For all amplitudes of quasiparticle transition the
antisymmetrization of the single quasiparticle states were
accounted for:\\
$\pi^f_4>\pi^f_3>\pi^f_2>\pi^f_4$,\\
$\eta^f_4>\eta^f_3>\eta^f_2>\eta^f_4$,\\
$\pi^i_4>\pi^i_3>\pi^i_2>\pi^i_4$,\\
$\eta^i_4>\eta^i_3>\eta^i_2>\eta^i_4$.\\
Further details of the formalism could be seen in Ref.\cite{Mut92}.

Nilsson model \cite{Nil55} was initially used to calculate wave
functions and single particle energies. Nuclear deformation was
taken into account in the Nilsson model.  We considered proton-neutron
residual interaction in two channels namely particle-particle
(attractive) and particle-hole (repulsive) interactions. In this
paper we emphasize that proton-neutron residual interaction is of
decisive importance in $\beta$-decay half-life calculation. A
reasonable choice for GT force parameters ($\chi$ and $\kappa$) may
lead to very good comparison of measured half-lives with the
calculated ones (e.g. Refs. \cite{Hir93,Sta90}). In the next section we would discuss this further. In present paper we
used the same range of values for the strength parameters as reported in
Ref. \cite{Maj17}. The value of $\chi$ = 61.20/A ($MeV$) and
$\kappa$ = 4.85/A ($MeV$) \cite{Maj17} were used in the current work.
The chosen values of $\chi$ and $\kappa$ display a $1/A$ dependence
as suggested in previous references \cite{Hom96,Nab15,Nab16,Nab17}.

The partial half-lives to daughter states ($E_f$) can be calculated
using the formula {
\begin{equation}\label{phl}
\begin{split}
t_{1/2}^{par} =\\
\frac {K}{(g_{A}/g_{V})^{2} f_{A}(Z,A,E)B_{GT}(E_{f}) +
f_{V}(Z,A,E)B_{F}(E_{f})}
\end{split}
\end{equation}

In Eq.~(\ref{phl}) $K$ is a constant of compound expressions taken
as 6143$s$ \cite{Har09}, $E = Q - E_f$ (where $Q$ is the energy released by the nuclear reaction), $g_A$ and $g_V$ are axial
and vector coupling constants, $f_A$ ($f_V$) is the Fermi integral
function for axial vector (vector) transitions. It is to be noted that the calculated Fermi integral function consists of two parts namely positron emission and electron capture and both contributions are taken into consideration in the current work. $B_F$ and
$B_{GT}$ are the reduced transition probabilities for the Fermi and
GT transitions, respectively. We can express these reduced
transition probabilities in the form of nuclear matrix elements as
follows

\begin{equation}
B_{GT} =\frac{1}{2I_{i}+1} \mid<f\parallel M_{GT} \parallel
i>\mid^{2} \label{rtp}
\end{equation}

and

\begin{equation}
B_{F} = \frac{1}{2I_{i}+1} \mid<f \parallel M_{F}
\parallel i> \mid ^{2}, \label{ftp}
\end{equation}
$M_{GT}$ represents GT transition  operator in Eq.~(\ref{rtp}) given
by
\begin{equation}
M_{GT} = \sum_{k} \tau_{\pm}(k) \sigma (k).
\end{equation}
The sum is carried over all nucleons in the nucleus. It is to be noted
that we only calculated $\tau_+$ $\sigma$ transitions in this work.
$M_{F}$ is the corresponding operator for Fermi transitions in
Eq.~(\ref{ftp}). Parent state spin is described by $I_{i}$ in these
equations.

The  deformation parameter ($\beta_2$) is another key parameter in
our nuclear model possessing deformed basis states. The value for
$\beta_2$ was taken from Ref. \cite{Mol16}. Q-values were taken from
the NUBASE2016 data \cite{Aud17}.

This paper further investigates pairing gaps effect on calculated
$\beta$-decay half-lives. Pairing gap computation was crucial for
the current calculation. For the calculation of pairing gaps, in
units of $MeV$, we used two different empirical formulae. The first
formula predicts different pairing gaps for protons and neutrons. It is a function of neutron separation energies ($S_n$) and
proton separation energies ($S_p$)(referred to as $Emp-1$ throughout this manuscript)
and shown in Eqs.~(\ref{PP}~-~\ref{NN})

\begin{equation}\label{PP}
\Delta_{pp}=
\frac{1}{4}(-1)^{Z+1}[S_{p}(A+1,Z+1)-2S_{p}(A,Z)+S_{p}(A-1,Z-1)]
\end{equation}

\begin{equation}\label{NN}
\Delta_{nn}=
\frac{1}{4}(-1)^{A-Z+1}[S_{n}(A+1,Z)-2S_{n}(A,Z)+S_{n}(A-1,Z)]
\end{equation}

The second formula for calculation of pairing gaps is the
traditionally used mass dependent recipe and same for protons and
neutrons (we refer to this formula as $Emp-2$ throughout this paper)
and given as

\begin{equation}\label{PG}
\Delta_{pp}=\Delta_{nn}=12/\sqrt{A}
\end{equation}

The $\beta$-decay half-life of a nucleus was obtained by summing up
all transition probabilities to states in the daughter nucleus with
excitation energies lying within the $Q_{\beta}$ window

\begin{equation}\label{T1-2}
T_{1/2} = (\sum_{0\le E_f \le Q_{\beta}}\frac
{1}{t_{1/2}^{par}})^{-1}.
\end{equation}

$E_f$ in Eq.~(\ref{T1-2}) represents daughter energy states and
$t_{1/2}^{par}$ are the partial half-lives introduced in
Eq.~(\ref{phl}).
\section{Results and Discussion}

Recently the extended QRPA (EQRPA) model, with and without
proton-neutron ($pn$) pairing, using two body interaction with
charge-dependent Bonn forces, was employed for calculation of
$\beta^{+}$/EC-decay half-lives of some medium mass neutron
deficient even-even isotopes of Cr, Fe, Ni, Zn, Ge and Se
\cite{Tan17}. We used  the pn-QRPA model, as discussed briefly in
former section, for calculation of $\beta$-decay half-lives for
selected Cr, Fe, Ni, Zn, Ge and Se isotopes as chosen in Ref.
\cite{Tan17}. These nuclei are important constituents of stellar
core of massive stars. All selected nuclei are even-even, medium
mass, $fp$-shell nuclei which plays key role in $rp$-process. It is
to be noted that our model could be employed for $\beta$-decay
half-life calculation of any arbitrary nucleus (and not
necessarily even-even nuclei).

We first investigate the impact of changing pairing strength
parameters on calculated GT strength distributions in our model. As
discussed earlier we  used the $Emp-1$ and $Emp-2$ scheme for
calculation of $\Delta_{pp} (\Delta_{nn})$ (see
Eqs.~(\ref{PP}~-~\ref{PG})).

Fig.~\ref{fig:1} depicts calculated GT strength
distributions for $^{46}$Cr, $^{50}$Fe and $^{54}$Ni using the two
schemes. It is noted that calculated GT strength distribution
changes appreciably for $^{46}$Cr and remains more or less the same
for $^{50}$Fe and $^{54}$Ni using the two different values of
pairing gaps. Similarly Fig.~\ref{fig:2} displays the corresponding
result for $^{62}$Zn, $^{66}$Ge and $^{70}$Se. The calculated
strength distributions do change appreciably with change in values
of $\Delta_{pp} (\Delta_{nn})$ (albeit less for the case of
$^{70}$Se).

Table~\ref{tab:1} shows the value of total strength (in
arbitrary units) and centroid  (in MeV units) of the
calculated GT strength distributions using different values of
pairing gaps for
all the nuclei considered in this paper. The last column mentions the cut-off energy in daughter up to which we show our calculation of total strengths and centroids (Columns V--VIII). It is noted from Table~\ref{tab:1} that we have up to a factor four difference between the calculated total GT strengths using the two different pairing gaps. We would like to comment that in most instances the GT strength redistributes itself when the pairing gaps change. They shift to energies higher than the cut-off energy shown in the last column of Table~\ref{tab:1}. This explains the, at times, large differences in calculated total strengths and centroids. This also has an effect on calculated half-lives which we discuss in next table.
It is clear from
Table~\ref{tab:1} that the pn-QRPA calculated GT strength
distribution is a sensitive function of pairing strength parameter.
The $S_p$ ($S_n$) dependent value of pairing strength ($Emp-1$)
results in calculated half-lives in good comparison with measured
data which we discuss next.

In Table~\ref{tab:2} we display the performance of our model
calculation. Shown are the pn-QRPA calculated half-lives along with
previous half-life calculations and measured half-lives for selected
even-even nuclei. The $Q$ values and experimental half-lives were
taken from Ref. \cite{Aud17}. Column IV and Column V show our
calculated half-lives using the $Emp-1$ and $Emp-2$ formulae for
pairing strength. The last three columns display previous
theoretical results. Column VI shows the calculated half-lives using
the FRDM model \cite{Mol97} whereas the last two columns show the
recent results of the EQRPA model \cite{Tan17}. Here we show both
EQRPA results  with and without the inclusion of proton-neutron
pairing correlations. All entries are given in units of $s$.
Comparison between measured data and pn-QRPA$^{(Emp-1)}$ shows
decent agreement. The calculated
root mean square (RMS) deviation from measured data for $Emp-1$, $Emp-2$, FRDM, EQRPA(no-pn) and EQRPA(with-pn) model calculations are 272.6$s$, 5488.1$s$, 1734.6$s$, 3869.3$s$ and 3229.3$s$, respectively. The low RMS value achieved from $Emp-1$ model is a clear indication of the fact that the predicted half-life values are in decent agreement with the measured data.
Noticeable differences in calculated pn-QRPA half-lives (using the
$Emp-1$ and $Emp-2$ formulae) are seen for six cases. For the cases
of $^{48}$Fe, $^{56}$Zn and $^{60}$Ge the $Emp-2$ calculated
half-lives are considerably bigger than those calculated using the
$Emp-1$ formula. The reason may be traced back to Table~\ref{tab:1}
where it is noted that total calculated GT strength using the
$Emp-1$ formula is appreciably bigger than the corresponding
strength using the $Emp-2$ formula. Bigger total strength translates
to bigger rates and correspondingly smaller half-lives. For the
nuclei $^{60,62}$Zn and $^{68}$Se the $Emp-2$ calculated half-lives
are appreciably smaller than those calculated using the $Emp-1$
formula. It is noted, from Table~\ref{tab:1}, that for all these
three cases the $Emp-2$ formula places the centroid of the GT
distribution at considerably lower daughter excitation energies than
those using the $Emp-1$ formula. This in turn translates into bigger
rates and corresponding smaller half-lives. In addition, for the
case of $^{60}$Zn and $^{68}$Se, the $Emp-1$ formula calculated
total GT strength is also much smaller than those calculated using
the $Emp-2$ formula. As a sample case, we present the state-by-state
half-life calculations of $^{62}$Zn in Table~\ref{tab:3} using the
$Emp-1$ and $Emp-2$ formulae for pairing strength. Shown also are
the pn-QRPA calculated GT strengths, branching ratios
$I_{(\beta^{+}/EC)}$ and partial half-lives $t^{par}_{1/2}$ for the
mentioned case. The branching ratio $I$ for each transition was
calculated using the following formula
\begin{equation}
I = \frac{T_{1/2}}{t^{par}_{1/2}}\times 100 (\%), \label{BR}
\end{equation}
where $T_{1/2}$ is the total half-life and $t^{par}_{1/2}$ is the
calculated partial half-life of the corresponding transition. It is noted from Table~\ref{tab:3} that it is the big branching ratio of 67$\%$ at 0.068 MeV in daughter $^{62}$Cu  that leads to a considerable decrease in the total calculated  half-life in the $Emp-2$ scheme.

It is evident that $\beta$-decay half-lives obtained as a result of
our $Emp-1$ scheme are in excellent agreement with the measured
half-lives. We achieve better agreement with measured data than the
previous calculations of Ref. \cite{Tan17} (using the EQRPA model)
and Ref. \cite{Mol97} (using the FRDM model).

As mentioned earlier tuning of GT force parameters ($\chi$ and $\kappa$)   can lead to good prediction power of the pn-QRPA model. In earlier calculations \cite{Sta90, Hir93}, instead of searching for a parameterization of GT force parameters, the authors directly applied the determined values of $\chi$ and $\kappa$ (using a procedure to match the measured half-lives). The predicted half-life values were nonetheless very encouraging. In previous pn-QRPA calculations \cite{Sta90, Hir93} it was argued that pairing gap values have very little affect on calculated $\beta$-decay half-lives. Since our finding is otherwise we decided to compare our calculated half-life values with previous pn-QRPA calculation \cite{Hir93}.   Table~\ref{tab:4} compares the current $Emp-1$ scheme calculation with the previously published half-life values from Ref. \cite{Hir93}. In Table~\ref{tab:4}, pn-QRPA(H) represents pn-QRPA calculated half-lives using mass formula of Ref. \cite{Hil76}. Similarly pn-QRPA(G) and pn-QRPA(M) represent pn-QRPA calculated half-lives using mass formula of Ref. \cite{Gro76} and \cite{Mol81}, respectively. We show comparison for only those nuclei where results were published earlier \cite{Hir93}. It may be noted that the half-lives calculated using the $Emp-1$ scheme are in good agreement with experimental half-lives. It is further noted that the pn-QRPA(M) predicted results are better than the  pn-QRPA(G) and pn-QRPA(H) predicted half-lives and in good agreement with measured data and $Emp-1$ scheme.

\section{Conclusions and Summary}
In this paper we calculated $\beta^{+}$/EC decay half-lives using the
pn-QRPA model for neutron deficient $fp$-shell nuclei. The
$\beta$-decay properties of chosen nuclei have a key role to play in
the nucleosynthesis problem.  It was concluded that the  pn-QRPA
model, in a multi-shell single-particle deformed space with
schematic interaction and reasonable choice of interaction constants
$\chi$ for particle-particle and $\kappa$ for particle-hole
parameters, results in accurate prediction of $\beta$-decay
half-lives. It was shown that pairing gaps  alter the
calculated GT strength distributions. It was further demonstrated
that $Emp-1$ formula for calculation of pairing gaps resulted in
better prediction of calculated half-life values than by using
$Emp-2$ scheme. Our calculated $\beta^{+}$/EC decay half-lives were in
excellent agreement with the measured one and showed marked
improvement over the former calculations. Because of the available
large model space (up to 7$\hbar \omega$ ) our model can calculate
the half-lives for any arbitrary heavy nucleus. It is expected that
the current investigation would lead to a better and reliable
calculation of $\beta$-decay properties of unstable nuclei. In future we wish to explore the effect of the two schemes ($Emp-1$  and  $Emp-2$) on predicted half-life values for the case of electron emission and double beta decay processes.

\section*{Acknowledgements}
J.-U. Nabi would like to acknowledge the support of the Higher
Education Commission Pakistan through project numbers 5557/KPK
/NRPU/R$\&$D/HEC/2016, 9-5(Ph-1-MG-7)/PAK-TURK /R$\&$D/HEC/2017 and
Pakistan Science Foundation through project number
PSF-TUBITAK/KP-GIKI (02).

M. Ishfaq wishes to acknowledge the support provided by Scientific
and Technological Research Council of Turkey (TUBITAK), Department
of Science Fellowships and Grant Programs (BIDEB) 2216 Research
Fellowship Program For International Researchers
(21514107-115.02-124287).

\begin{table}[htbp]
\center \scriptsize\caption{\scriptsize Comparison between pn-QRPA
calculated Gamow-Teller strength distributions using different
values of pairing gaps: $\Delta_{nn}^{(Emp-1)}$ ,
$\Delta_{pp}^{(Emp-1)}$ and $\Delta_{nn,pp}^{(Emp-2)}$. Shown are
the calculated pairing gaps (in $MeV$), total Gamow-Teller strength,
centroid and cut-off energy (in $MeV$) using the
pn-QRPA($Emp-1$) and pn-QRPA($Emp-2$) scheme for selected nuclei.}
\label{tab:1} \hspace{0.1in}
\begin{tabular}{llllllllllll}
\hline\noalign{\smallskip}
Nucleus & $\Delta_{nn}^{(Emp-1)}$ & $\Delta_{pp}^{(Emp-1)}$ & $\Delta_{nn,pp}^{(Emp-2)}$ & $\sum GT^{Emp-1}$  & $\sum GT^{Emp-2}$   & ${\bar E^{Emp-1}}$ & ${\bar E^{Emp-2}}$ & E$_{cut off}$\\
\noalign{\smallskip}\hline\noalign{\smallskip}
$^{42}$Cr&1.78 &1.97 &1.85  &11.6 & 4.58 & 7.40 & 5.69 & 13.8  \\
$^{44}$Cr&2.07 &1.78 &1.81  &1.66 & 1.69 & 0.45 & 0.45 & 13.9  \\
$^{46}$Cr&2.24 &1.94 &1.77  &1.63 & 1.72 & 0.43 & 0.43 & 8.0 \\
$^{46}$Fe&3.04 &1.79 &1.77  &10.7 & 10.9 & 6.43 & 7.03 & 13.5  \\
$^{48}$Fe&1.79 &1.67 &1.73  &2.42 & 0.78 & 1.51 & 4.59 & 13.5  \\
$^{50}$Fe&1.86 &1.51 &1.70  &26.6 & 26.6 & 4.37 & 4.22 & 8.1  \\
$^{48}$Ni&1.87 &0.66 &1.73  &40.3 & 14.0 & 8.18 & 9.93 & 15.6  \\
$^{50}$Ni&2.09 &1.86 &1.70  &10.4 & 2.89 & 6.98 & 5.39 & 12.7  \\
$^{52}$Ni&2.02 &1.69 &1.66  &7.97 & 5.23 & 3.92 & 2.92 & 11.0  \\
$^{54}$Ni&1.48 &1.64 &1.62 &1.71 & 1.71 & 0.002 & 0.002& 11.0  \\
$^{56}$Zn&1.71 &1.35 &1.60  &14.4 & 3.56 & 9.94 & 7.03 & 12.3  \\
$^{58}$Zn&1.90 &1.23 &1.58  &11.3 & 25.1 & 5.30 & 4.68 & 9.2  \\
$^{60}$Zn&1.71 &1.64 &1.55  &1.83 & 2.78 & 2.83 & 2.17 & 4.5  \\
$^{62}$Zn&1.60 &1.37 &1.52  &0.98 & 0.51 & 1.10 & 0.06 & 1.6  \\
$^{60}$Ge&1.97 &1.41 &1.55  &14.4 & 4.00 & 9.30 & 8.19 & 12.2  \\
$^{62}$Ge&1.30 &1.21 &1.52  &13.7 & 10.7 & 6.90 & 7.25 & 10.0  \\
$^{64}$Ge&1.90 &1.88 &1.50  &0.58 & 0.50 & 1.55 & 1.58 & 4.5  \\
$^{66}$Ge&1.76 &3.47 &1.48  &0.58 & 0.66 & 1.05 & 1.26 & 2.0  \\
$^{64}$Se&1.60 &1.42 &1.50  &15.9 & 14.3 & 6.87 & 7.20 & 12.7  \\
$^{66}$Se&1.55 &1.44 &1.48  &14.7 & 6.57 & 4.36 & 3.15 & 10.6  \\
$^{68}$Se&1.94 &2.08 &1.46  &3.85 & 8.99 & 3.18 & 0.57 & 4.7  \\
$^{70}$Se&1.88 &1.73 &1.43  &0.99 & 1.05 & 1.24 & 1.19 & 2.4  \\
\noalign{\smallskip}\hline
\end{tabular}
\end{table}

\clearpage
\begin{table}[htbp]
\center\scriptsize \caption{\scriptsize Comparison between
experimental \cite{Aud17}, pn-QRPA$^{(Emp-1)}$, pn-QRPA$^{(Emp-2)}$,
FRDM model \cite{Mol97} and EQRPA model (with and without $pn$
pairing) \cite{Tan17} calculated half-lives. Q$_{\beta}$ values were
taken from Ref. \cite{Aud17} and given in units of $MeV$. All
half-lives are given in units of $s$.} \label{tab:2} \hspace{0.1in}
\begin{tabular}{llllllll}
\hline\noalign{\smallskip}
Nucleus & $Q_{\beta}$  & $T^{[EXP]}_{1/2}$ & $T^{[pn-QRPA(Emp-1)]}_{1/2}$  & $T^{[pn-QRPA(Emp-2)]}_{1/2}$ & $T^{[FRDM]}_{1/2}$  & $T^{[EQRPA(no-pn)]}_{1/2}$ & $T^{[EQRPA(with-pn)]}_{1/2}$ \\
\noalign{\smallskip}\hline\noalign{\smallskip}
$^{42}$Cr& 13.7 & 0.013 & 0.013  & 0.013  & 0.045 & 0.013 & 0.012  \\
$^{44}$Cr& 10.5 & 0.042 & 0.038  & 0.037  & 0.118 & 0.056 & 0.038  \\
$^{46}$Cr& 7.60 & 0.224 & 0.218  & 0.209  & 0.671 & 0.404 & 0.375 \\
$^{46}$Fe& 13.5 & 0.013 & 0.012  & 0.012  & 0.018 & 0.014 & 0.011\\
$^{48}$Fe& 10.9 & 0.045 & 0.042  & 1.123  & 0.059 & 0.037 & 0.034\\
$^{50}$Fe& 8.14 & 0.152 & 0.135  & 0.119  & 0.542 & 0.301 & 0.301\\
$^{48}$Ni& 15.6 & 0.003 & 0.003  & 0.003  & 0.005 & 0.005 & 0.002\\
$^{50}$Ni& 12.9 & 0.018 & 0.018  & 0.018  & 0.017 & 0.018 & 0.016\\
$^{52}$Ni& 10.5 & 0.041 & 0.039  & 0.039  & 0.077 & 0.056 & 0.052\\
$^{54}$Ni& 8.79 & 0.114 & 0.102  & 0.102  & 0.646 & 0.329 & 0.299\\
$^{56}$Zn& 12.7 & 0.032 & 0.029  & 0.285  & 0.083 & 0.025 & 0.021\\
$^{58}$Zn& 9.37 & 0.086 & 0.048  & 0.043  & 0.597 & 0.192 & 0.162\\
$^{60}$Zn& 4.17 & 142.8 & 142.5  & 37.33  & $>$100& 268.3 & 60.29\\
$^{62}$Zn& 1.62 & 33094 & 32101  & 7399   & $>$100& 39372 & 31372\\
$^{60}$Ge& 12.2 & 0.030 & 0.025  & 0.245  & 0.082 & 0.494 & 0.424\\
$^{62}$Ge& 10.1 & 0.129 & 0.117  & 0.081  & 0.868 & 0.125 & 0.102\\
$^{64}$Ge& 4.52 & 63.70 & 48.62  & 50.16  & 80.88 & 752.2 & 665.6\\
$^{66}$Ge& 2.12 & 8136  & 7452   & 9557   & $>$100& 25125 & 23144\\
$^{64}$Se& 12.7 & 0.030 & 0.028  & 0.030  & 0.097 & 0.022 & 0.020\\
$^{66}$Se& 10.7 & 0.030 & 0.031  & 0.033  & 0.648 & 0.073 & 0.065\\
$^{68}$Se& 4.71 & 35.50 & 35.06  & 0.759  & 42.32 & 17.69 & 17.68\\
$^{70}$Se& 2.41 & 2466  & 2041   & 1865   & $>$100 & 3388 & 3387\\
\noalign{\smallskip}\hline
\end{tabular}
\end{table}
\clearpage

\clearpage
\begin{table}[htbp]
\center\scriptsize \caption{\scriptsize The state-by-state half-life
calculations for $^{62}$Zn using the Emp - 1 and Emp - 2 formulae
for pairing strength. $E_{x}$ are daughter excited state energies in
units of $MeV$. Shown also are the corresponding GT
strengths, branching ratios $I_{(\beta^{+}/EC)}$ and partial
half-lives. All half-lives are given in units of $s$.} \label{tab:3}
\hspace{0.1in}
\begin{tabular}{llll|llll}
\hline\noalign{\smallskip}
$E_{x}^{Emp-1}$ & $GT^{Emp-1}$ & $I_{\beta^{+}/EC}$ [Emp-1] & $t^{par}_{1/2}$[Emp-1] & $E_{x}^{Emp-2}$  & $GT^{Emp-2}$  & $I_{\beta^{+}/EC}$ [Emp-2] & $t^{par}_{1/2}$[Emp-2] \\
\noalign{\smallskip}\hline\noalign{\smallskip}
0.206& 0.01297& 8.1880& 3.92E+05& 0.000& 0.14427& 32.849& 2.25E+04 \\
0.208& 0.02066& 12.997& 2.46E+05& 0.068& 0.34368& 66.717& 1.10E+04\\
0.259& 0.01591& 9.0970& 3.52E+05& 0.921& 0.01203& 0.4000& 1.84E+06\\
0.601& 0.00384& 1.1870& 2.70E+06& 1.194& 0.00076& 0.0090& 7.99E+07\\
0.654& 0.01223& 3.3930& 9.46E+05& 1.194& 0.00009& 0.0010& 6.56E+08\\
0.758& 0.02285& 5.0370& 6.37E+05& 1.325& 0.00047& 0.0030& 2.75E+08\\
0.953& 0.04047& 5.3110& 6.04E+05& 1.340& 0.00407& 0.0210& 3.52E+07\\
0.957& 0.07971& 10.339& 3.10E+05&- &- &- & -\\
1.003& 0.09434& 10.576& 3.03E+05&- &- &- & -\\
1.158& 0.12661& 7.8700& 4.07E+05&- &- &- & -\\
1.164& 0.42584& 25.823& 1.24E+05&- &- &- & -\\
1.534& 0.07499& 0.1370& 2.34E+07&- &- &- & -\\
1.554& 0.04646& 0.0460& 7.02E+07&- &- &- & -\\
1.588& 0.00259& 0.0000& 7.60E+09&- &- &- & -\\
\noalign{\smallskip}\hline
\end{tabular}
\end{table}

\clearpage
\begin{table}[htbp]
\center\scriptsize \caption{\scriptsize Comparison of present pn-QRPA$^{(Emp-1)}$ and previous pn-QRPA calculation \cite{Hir93} with measured half-lives
\cite{Aud17}.  All half-lives are given in units of $s$.}
\label{tab:4} \hspace{0.1in}
\begin{tabular}{llllll}
\hline\noalign{\smallskip}
Nucleus & $T^{[EXP]}_{1/2}$ & $T^{[pn-QRPA(Emp-1)]}_{1/2}$  & $T^{[pn-QRPA(H)]}_{1/2}$ & $T^{[pn-QRPA(G)]}_{1/2}$  & $T^{[pn-QRPA(M)]}_{1/2}$ \\
\noalign{\smallskip}\hline\noalign{\smallskip}
$^{42}$Cr& 0.013 & 0.013  & 0.033  & 0.025 & 0.012 \\
$^{44}$Cr& 0.042 & 0.038  & 0.083  & 0.068 & 0.041 \\
$^{46}$Fe& 0.013 & 0.013  & 0.031  & 0.024 & 0.015 \\
$^{48}$Ni& 0.003 & 0.003  & 0.011  & 0.008 & 0.006 \\
$^{50}$Ni& 0.018 & 0.018  & 0.018  & 0.014 & 0.015 \\
$^{60}$Ge& 0.030 & 0.025  & 0.034  & 0.031 & 0.032 \\
$^{62}$Ge& 0.129 & 0.117  & 0.086  & 0.082 & 0.083 \\
$^{64}$Se& 0.030 & 0.028  & 0.033  & 0.032 & 0.030 \\
$^{66}$Se& 0.030 & 0.031  & 0.073  & 0.072 & 0.071 \\
\noalign{\smallskip}\hline
\end{tabular}
\end{table}

\clearpage
\begin{figure}[p]
\begin{center}
  \includegraphics[width=1.0\textwidth]{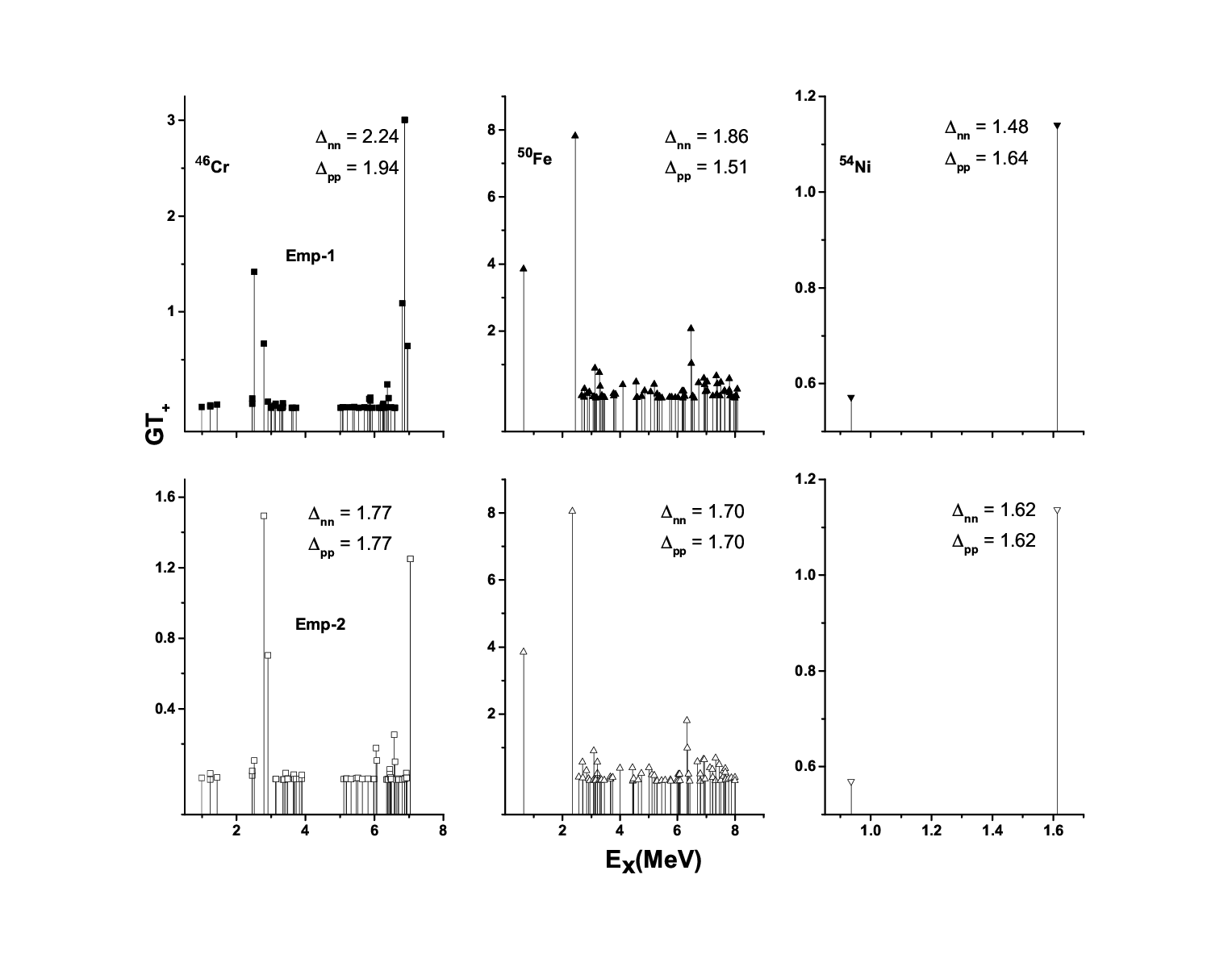}
 \caption{\scriptsize Calculated GT strength distributions for $^{46}$Cr,
$^{50}$Fe and $^{54}$Ni using pn-QRPA$^{(Emp-1)}$ and
pn-QRPA$^{(Emp-2)}$ schemes.} \label{fig:1}
\end{center}
\end{figure}

\clearpage
\begin{figure}[p]
\begin{center}
  \includegraphics[width=1.0\textwidth]{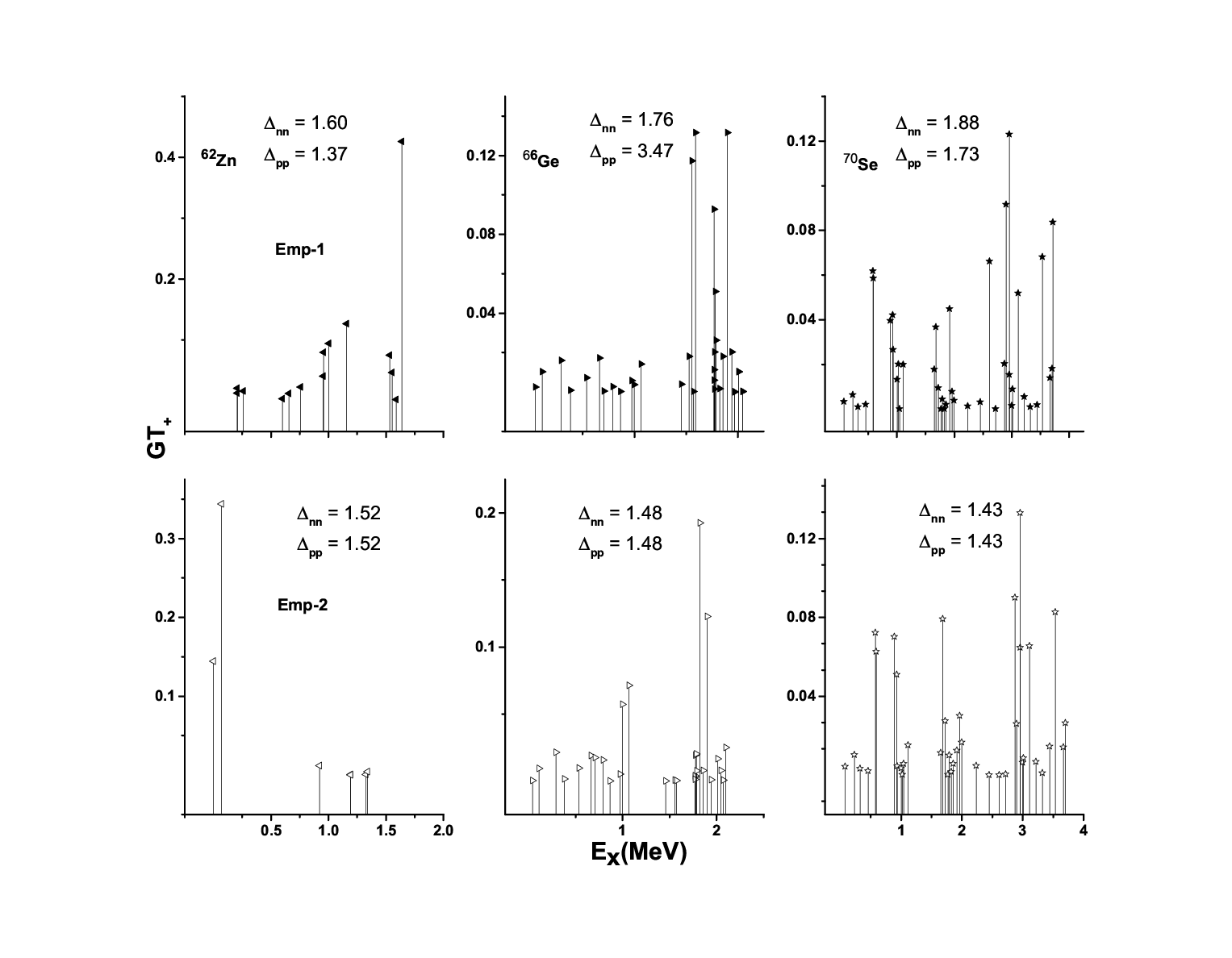}
\caption{\scriptsize Calculated Gamow-Teller strength distributions
for $^{62}$Zn, $^{66}$Ge and $^{70}$Se using pn-QRPA$^{(Emp-1)}$ and
pn-QRPA$^{(Emp-2)}$ schemes.} \label{fig:2}
\end{center}
\end{figure}

\end{document}